\documentclass[a4paper,10pt,twoside]{cpc-hepnp}
\usepackage{color}
\usepackage{multicol}
\usepackage{graphicx}
\usepackage{booktabs}
\usepackage{amssymb,bm,mathrsfs,bbm,amscd}
\usepackage[tbtags]{amsmath}
\usepackage{lastpage}

\begin{document}

\fancyhead[c]{\small Chinese Physics C~~~Vol. XX, No.XX (201X) XXXXXX}
\fancyfoot[C]{\small 010201-\thepage}

\footnotetext[0]{Received XX XX 201X}

\title{Potential of Geo-neutrino Measurements at JUNO\thanks{Supported in part by National Natural Science
Foundation of China (11405056, 11305193, 11205176, 21504063), by National Science Foundation of U.S. (EAR 1067983/1068097),
by the Strategic Priority Research Program of the Chinese Academy of Sciences under Grant No. XDA10010100
and by the CAS Center for Excellence in Particle Physics (CCEPP) }}

\author{%
Ran HAN$^{1;1)}$\email{han.ran@gmail.com}%
\quad Yu-Feng LI$^{2;2)}$\email{liyufeng@ihep.ac.cn}%
\quad Liang ZHAN$^{2}$
\quad William F McDonough$^{3}$
\quad Jun CAO$^{2}$
\quad Livia Ludhova$^{4}$
}

\maketitle

\address{%
$^1$ Science and Technology on Reliability and Environmental Engineering Laboratory, Beijing Institute of Spacecraft Environment Engineering, Beijing 100094, China\\
$^2$ Institute of High Energy Physics, Chinese Academy of Sciences, Beijing 100049, China\\
$^3$ Department of Geology, University of Maryland, College Park, MD, USA 20742\\
$^4$ Forschungzentrum J\"ulich IKP-2 and RWTH Aachen University, Germany
}

\begin{abstract}
The flux of geoneutrinos at any point on the Earth is a function of the abundance and distribution of radioactive elements within our planet. This flux has been successfully detected by the 1-kt KamLAND and 0.3-kt Borexino detectors with these measurements being limited by their low statistics. The planned 20-kt JUNO detector will provide an exciting opportunity to obtain a high statistics measurement, which will provide data to address several questions of geological importance. This paper presents the JUNO detector design concept, the expected geo-neutrino signal and corresponding backgrounds. The precision level of geo-neutrino measurements at JUNO is obtained with the standard least-squares method. The potential of the Th/U ratio and mantle measurements is also discussed.
\end{abstract}

\begin{keyword}
Geoneutrinos; JUNO; Precision, Th/U ratio, Mantle
\end{keyword}

\begin{pacs}
1---3 PACS(Physics and Astronomy Classification Scheme, http://www.aip.org/pacs/pacs.html/)
\end{pacs}

\footnotetext[0]{\hspace*{-3mm}\raisebox{0.3ex}{$\scriptstyle\copyright$}2013
Chinese Physical Society and the Institute of High Energy Physics
of the Chinese Academy of Sciences and the Institute
of Modern Physics of the Chinese Academy of Sciences and IOP Publishing Ltd}%

\begin{multicols}{2}

\section{Introduction}

Some 150 years ago Lord Kelvin brought to prominence the discussion concerning the age of the Earth and he linked it to its cooling history. He assumed a simple model of solid-state cooling of a sphere that did not convect and contained no internal heating source. Today we understand that cooling of the core and the mantle is controlled by convective heat transfer and that radioactive elements (i.e., potassium, thorium and uranium, K, Th and U, the heat producing elements, HPE) contribute to the Earth's surface heat flux. The Earth's surface heat flow $46\pm3$ TW~\cite{Jaupart:2007} has been firmly established for the last half century, however, vigorous debate continues regarding the relative contributions of primordial versus radioactive sources. At the fundamental level this debate relates to the composition of the Earth, the distribution of HPE, and whether or not there is chemical layering in the mantle, which in turn relates to the nature and form of convection in the mantle. Measuring the Earth's geoneutrino flux, electron anti-neutrinos produced during the beta-minus decay of HPE, will provide insights into defining the power driving mantle convection, plate tectonics and geodynamo, with the latter producing the magnetosphere that shields the Earth from harmful cosmic ray flux.

Over the last decade particle physicists have detected the Earth's geoneutrino flux~\cite{McDonough:2012zz}. Neutrinos and their anti-particle counter-parts are nearly massless and uncharged elementary particles that travel at close to the speed of light. Matter, including the Earth, is mostly transparent to these elusive messengers as they virtually escape detection. The Earth produces an electron anti-neutrino flux of $\sim 6$ million per centimeter squared per second and by detecting a few of these particles per year we are now measuring the thorium and uranium content inside the planet, which in turn allows us to determine the amount of radiogenic power driving the Earth's engine.

Detection of geo-neutrinos serves the geology, particle physics, and nuclear security communities collectively and separately. The particle physics community seeks to understand the nature of these particles, whereas the nuclear security community needs to understand the flux from active nuclear reactors, which requires that they accurately subtract the overlapping geoneutrino background signal. The geological community seeks to have transformational insights from their detection by accurately determining the global inventory of HPE. The combination of K, Th and U account for more than 99\% of the radiogenic heat production in the Earth and together with the primordial energy of accretion and core segregation define the total power budget of the planet. Current and future liquid-scintillator detectors will directly measure the amount of Th and U in the Earth, but not the signal from K, because the geo-neutrino energy from the beta decay of K is less than that needed to initiate the inverse beta decay mechanism used to record events in these detectors.

The structure of the Earth and the distribution of HPE will be briefly discussed in the following.
The Earth is made up of two layers, the metallic core and the silicate Earth. The core is internally differentiated into a small ($\sim 5\%$ by mass) solid inner core and a much larger liquid outer core, both of which are considered to contain negligible quantities of HPE. The silicate Earth, the host of the planet's HPE budget, is composed of the mantle and crust, with the latter being differentiated into basaltic oceanic crust, which has a low content of HPE, and granitic-like continental crust, which has a factor of 10 or more greater HPE content. The mantle of the Earth makes up 2/3 of the planet's mass and its structure, which covers half the planetary radius, is still poorly understood in terms of chemical layering, convection, and the fate of seafloor fragments being transferred back into the mantle via processes of plate tectonics at deep sea trenches. Our knowledge of the HPE content in the oceanic and continental crusts is considerably more mature, given it is the accessible Earth. Based on our knowledge of the Earth's crusts and competing compositional models of the Earth~\cite{McDonough:1995}, we are left with a factor of 30 uncertainty in the composition of the mantle for its content of HPE~\cite{Sramek:2012nk}.

The particle physics community is taking a bold new approach to the detection of electron anti-neutrinos with the development of the JUNO detector sited in southern China.  The JUNO detector, which stands for the Jiangmen Underground Neutrino Observatory~\cite{Djurcic:2015vqa,An:2015jdp}, is a 20 kton liquid-scintillator experiment, comparable to, but 20 times greater than the existing Japanese KamLAND detector (the first experiment to measure the Earth's geoneutrino flux)~\cite{Araki:2005qa} and the soon to come online Canadian SNO+ detector~\cite{Chen:2005zza}. Moreover, JUNO is 60 times more massive than the existing Italian Borexino experiment~\cite{Bellini:2010hy}. Thus, the new JUNO detector, which will have a significant reactor signal as the experiment is dedicated to fundamental studies of these particles and their properties, will collect a large geo-neutrino signal annually and thus have the potential to inform the geology community about the Earth's total flux and details of the contribution from the region surrounding the detector.

The JUNO detector, sited near the southern coast of China, presents an opportunity to the geology community to detect the Earth's geo-neutrino signal along a continental margin, a first for this setting. The detected signal, which can be decomposed into contributions from the continental crust, the oceanic crust, and the mantle, will be compared with those detected at other existing experiments to extract the mantle signal and reveal the crustal contribution.
Given the large size, the JUNO detector will rapidly accumulate its geo-neutrino signal. Within about one year it will have more
geo-neutrino events than all detectors combined will have accumulated to that time.
Therefore, the JUNO detector has the potential to greatly contribute to our understanding of the Earth. However, the setting for the JUNO experiment,
which is strategically placed 53 km from two large nuclear power plants to optimize the physics experimental goals, poses a large challenge for extracting the geo-neutrino signal. This paper will address the strengths and weaknesses of the JUNO experiment for geo-neutrino detection. It will introduce the design of the detector and experimental facility, identify the expected geo-neutrino signal at JUNO, and it will address the issue of identifying and subtracting the reactor and other background signals from the geo-neutrino signal. The standard least-squares method will be used to quantitatively evaluate the potential of the geo-neutrino measurement.

\section{Geo-neutrino signal at JUNO}

Located at Kaiping, Jiangmen, in South China, the JUNO detector is about 53 km from the Yangjiang and Taishan nuclear power plants, which in total will have thermal power of 36 GW. The JUNO experiment is designed to determine the neutrino mass hierarchy and precisely measure oscillation parameters by detecting reactor antineutrinos from nuclear power plants. The high-purity, liquid-scintillator detector is 20 kt in size and will be sited 700 meters underground to shield it from cosmic ray fluxes. The large detector size and ultra-clean experimental environment will allow JUNO to have the possibility of seeing supernova neutrinos, along with studying atmospheric, solar and geo-neutrinos.

The design of the central detector envisages a spherical acrylic tank surrounded by a water-based muon veto layer. The acrylic tank will be filled with 20 kt of linear alkylbenzene (LAB) liquid scintillator (LS). Facing the scintillation volume will be 17,000 20-inch photomultiplier tubes (PMTs) installed on a surrounding stainless steel truss. The enveloping outer water pool protects the central detector from natural radioactivity in surrounding rocks and PMTs. It also serves as a water Cherenkov detector after being equipped with PMTs, to identify and track cosmic muons. A second muon tracking detector on top of the water pool is used to improve muon detection efficiency and tracking.

The expected geoneutrino signal at JUNO was calculated in Ref.~\cite{Strati:2014kaa}, where they adopted the reference Earth model (RM) developed by Huang et al~\cite{Huang:2013}. The RM model divides the silicate Earth into eight lithospheric reservoirs (ice, water, soft and hard sediment, upper, middle and lower continental crust, and lithospheric mantle) and two mantle reservoirs (depleted and enriched). The model uses a resolution of $1^\circ \times 1^\circ$ for setting the abundance of Th and U in each layer based on geochemical and geophysical inputs. The geo-neutrino flux at any detector site is calculated using a detailed regional lithospheric model (the closest 500 km from the detector site) that is coupled to a global, far-field model of the lithosphere and the underlying mantle models. Strati {\it et al.}~\cite{Strati:2014kaa} estimated a total geo-neutrino signal from U in the lithosphere of $23.2^{+5.9}_{-4.8}$ TNU (Terrestrial Neutrino Unit,
1 TNU = 1 events/year/$10^{32}$ protons). The asymmetric $1\sigma$ errors are obtained from Monte Carlo simulations following the method of Huang {\it et al.}~\cite{Huang:2013} and account only for uncertainties from the lithosphere. From this signal of U we can then calculate the Th contribution based on an assume chondritic Th/U ratio. Given the abundance of U and Th and mass of a given layer in the Earth, the radiogenic heat production can be calculated as~\cite{Fiorentini:2007te}:
\begin{equation}
\label{eq:HRThU}
H_R({\rm U+Th})=9.85\times m({\rm U})+2.67\times m({\rm Th}).
\end{equation}
The total uranium signal, as a function of the uranium mass contained in the lithosphere $m_L({\rm U})$ and in the mantle $m_M({\rm U})$, will be
\begin{eqnarray}
\label{eq:SThU}
S({\rm U})&=&S_L({\rm U})+S_M({\rm U})=S_L({\rm U})+\beta \times m_M({\rm U})\nonumber\\
&=&S_L({\rm U})+\beta \times [m({\rm U})-m_L({\rm U})],
\end{eqnarray}
where $S_L({\rm U})=23.2^{+5.9}_{-4.8}$ TNU~\cite{Strati:2014kaa} is the signal from lithosphere and $\beta$ defines the signal contribution from the mantle.
We assume two limiting cases for the value of $\beta$ in a spherically symmetric mantle:
\begin{eqnarray}
S_{M,low}({\rm U})=\beta_{low}\times [m({\rm U})-m_{L,high}({\rm U})] \quad{\rm TNU},\\
S_{M,high}({\rm U})=\beta_{high}\times [m({\rm U})-m_{L,low}({\rm U})] \quad {\rm TNU},
\end{eqnarray}
with $\beta_{low}=12.15$ TNU and $\beta_{high}=17.37$ TNU, which correspond to placing
placing radioactive uranium in a thin layer at the bottom and uniformly over the mantle, respectively.
The mass $m_L({\rm U})$ is given in the unit of $10^{17}$ kg. $m_{L,high}({\rm U})=0.4$ and $m_{L,low}({\rm U})=0.3$ are two limiting values of the
U mass in the lithosphere. 
The Th and U contributions in the chondritic proportion i.e., $m({\rm Th})/m({\rm U}) = 3.9$, combining with Eqs.~(\ref{eq:HRThU}) and (\ref{eq:SThU}), give the expected total geo-neutrinos signal $S(\rm U+Th)$ at JUNO as a function of radiogenic heat $H(\rm U+Th)$, which is depicted in Fig.~\ref{fig:SH}. The red and blue lines correspond to the assumed values of $\beta_{high}$ and $\beta_{low}$.
From left to right, the parallelogram region between the red and blue lines denotes the region allowed by cosmochemical, geochemical and geodynamical models, respectively~\cite{Sramek:2012nk}.
\begin{center}
\includegraphics[width=8.5cm]{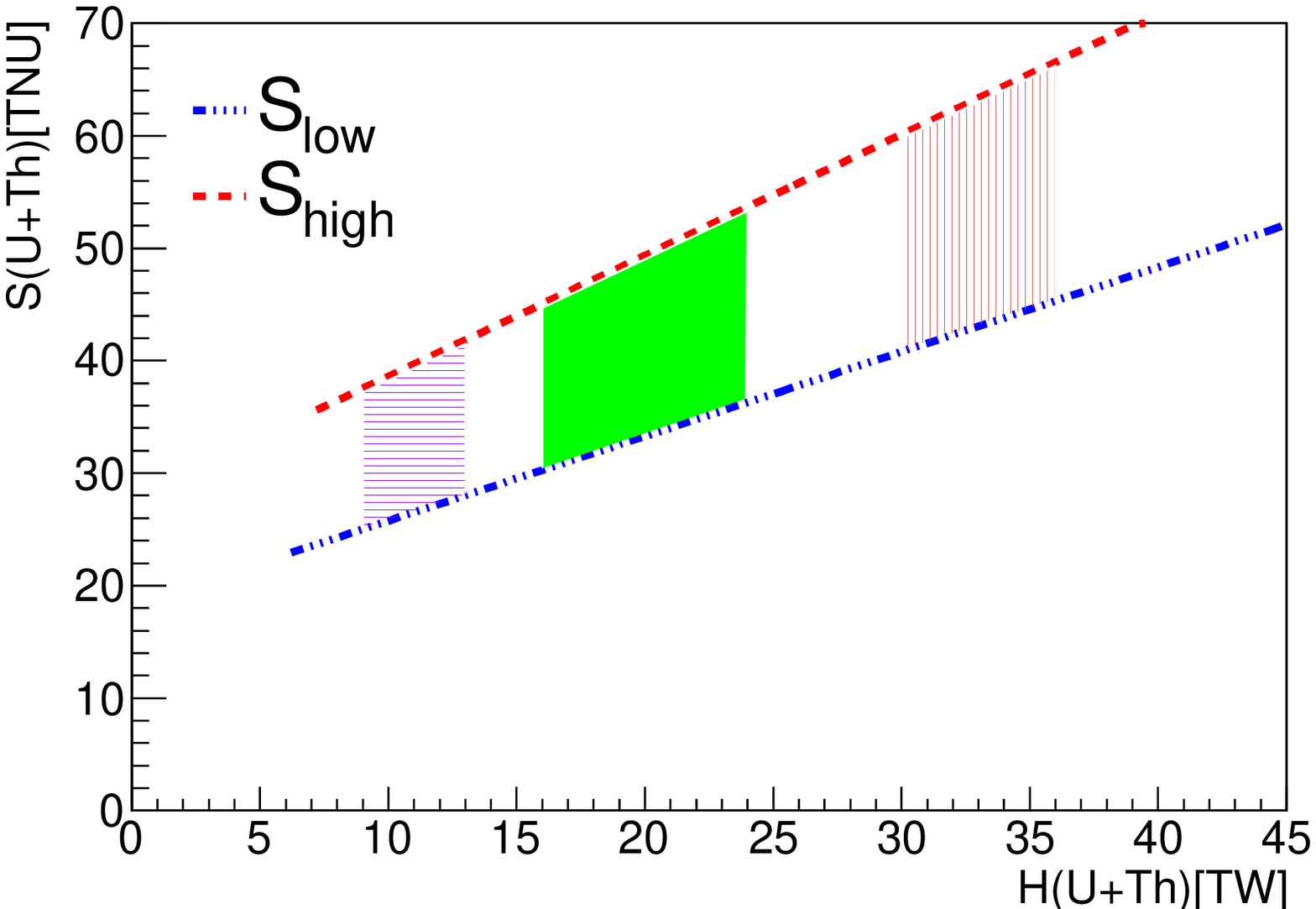}
\vspace{1mm}
\figcaption{\label{fig:SH} The expected geoneutrino signal at JUNO as a function of radiogenic heat due to U and Th in the Earth $H(\rm U + Th)$.
The red line and blue line correspond to two limiting cases .
From left to right, the quadrangle regions between the red and blue lines denote the regions allowed by cosmochemical,geochemical and geodynamical models, respectively~\cite{Sramek:2012nk}.}
\end{center}

\section{Backgrounds}

Compared with the KamLAND (Japan) and Borexino (Italy) experiments, JUNO has a much larger detector volume and can collect more geo-neutrino events. However, JUNO suffers a larger background from reactor antineutrinos, generated by the Yangjiang and Taishan nuclear power plants. Other non-antineutrino backgrounds may also be a challenge for geo-neutrino detection.
\subsection{Reactor antineutrinos}
In reactors, electron antineutrinos are emitted mainly from the fissions of four isotopes, $^{235}\rm U$, $^{238}\rm U$, $^{239}\rm Pu$, and $^{241}\rm Pu$. The expected antineutrino spectrum at a reactor is predicted as:
\begin{eqnarray}
\label{eq:reactorflux}
\Phi(E_{\bar{\nu}_e})=\frac{W_{th}}{\sum_i{f_i}\cdot{Q_i} }\sum_i f_{i} \cdot S_{i}(E_{\bar{\nu}_e}),
\end{eqnarray}
where $f_i$ and $Q_i$ are the fission fraction and the energy released per fission of the $i$-th isotope. $S_i$ is the antineutrino spectrum of the $i$-th isotope.
$W_{th}$ is the reactor thermal power.
The fission fractions are taken as $\mbox{$^{235}$U} : \mbox{$^{238}$U} : \mbox{$^{239}$Pu}  : \mbox{$^{241}$Pu}   =
0.577 :  0.076 : 0.295 : 0.052 $,
which are the average values of the Daya Bay reactor cores and represent the typical values of the PWR and BWR reactors.
The energy spectra are taken from Ref.~\cite{Mueller:2011nm}, and the energy released per fission of the $i$-th component is from Ref.~\cite{Ma:2012bm}.

The reactor antineutrinos are detected by the inverse beta decay (IBD) reaction.
The expected number of reactor antineutrino events in the detector is predicted as:
\begin{eqnarray}
\label{eq:reactorspec}
&&N_{\bar{\nu}_e}= \epsilon \times {N_p} \times \tau \times\nonumber\\
&&\sum_{r=1}^{N_{\rm rea}} \frac{1}{4\pi {L_r}^2} \int dE_{\bar{\nu}_e} \sigma(E_{\bar{\nu}_e}) P_{ee}(E_{\bar{\nu}_e}, L_r) \Phi(E_{\bar{\nu}_e}),
\end{eqnarray}
where $\epsilon$ is the detector efficiency corresponding to a IBD selection criteria, $N_p$ is the number of free protons and $\tau$ is the data-taking time.
The index $r$ runs over the number of reactors, $L_r$ is the baseline from the detector to the reactor.
The $\sigma(E_{\bar{\nu}_e})$ is the cross section of the IBD reaction, $P_{ee}(E_{\bar{\nu}_e}, L_r)$ is the electron antineutrino survival probability and $\Phi(E_{\bar{\nu}_e})$ is the expected reactor antineutrino spectrum from Eq.~(\ref{eq:reactorflux}).

To estimate the event number of reactor antineutrinos at JUNO, we first consider the contribution from all the reactor cores in the world in operation
in 2013. The contribution of the reactor cores in operation in 2013 is taken from Ref.~\cite{Baldoncini:2014vda}, which gives $95.3^{+2.6}_{-2.4}$ TNU.

The contribution of Taishan and Yangjiang nuclear power plants are estimated using Eq.~(\ref{eq:reactorspec}). The IBD detection efficiency is assumed to be 80\% and the fiducial volume is 18.35~kton with a 17.2~m radial cut, which yields $1.285 \times 10^{33}$ free protons. The thermal power $W_{\rm th}$ and baseline $L_r$ for each reactor core in Taishan and Yangjiang are taken from Ref.~\cite{Li:2013zyd} and the IBD cross section $\sigma(E_{\bar{\nu}_e})$ is taken from Ref.~\cite{Strumia:2003zx}. The oscillation parameters in the survival probability $P_{ee}(E_{\bar{\nu}_e}, L_r)$ are taken from Ref.~\cite{Capozzi:2013csa}.

Eq.~(\ref{eq:reactorspec}) can also be used to estimate the uncertainties of the reactor IBD background.
The correlated uncertainties between reactors include those from the energy per fission
($0.2\%$) and the IBD reaction rate ($2.7\%$). The uncorrelated uncertainties between reactors include the thermal power ($0.5\%$),
the fission fraction ($0.6\%$), non-equilibrium effects ($0.3\%$) and the contribution from spent nuclear
fuel ($0.3\%$). The uncertainty from oscillation parameters is mainly from $\theta_{12}$, which is estimated to
be negligible, considering a sub-percent level can be obtained with the JUNO detector itself.
As a result, the total uncertainty is $2.8\%$.

In summary, the expected reactor antineutrino events from all nuclear cores in the world operating in 2013 is $980^{+27}_{-25}$ events per year and the contribution from Taishan and Yangjiang nuclear power plants is $15120\pm423$ events per year.

\subsection{Non-antineutrino backgrounds}

In addition to the reactor antineutrino background, there are other non-antineutrino backgrounds relevant for geo-neutrino detection.

The $\beta$-n decays from $^{9}{\rm Li}$ and $^{8}{\rm He}$ isotopes produced by cosmic muons crossing the detector can mimic IBD reactions. The total rate of $\beta$-n decays is 84/day in the whole central detector. However, this $\beta$-n background can be effectively reduced using muon veto criteria, which employ both the time and space distribution of isotope products with respect to their tagged mother muons (See Ref.~\cite{An:2015jdp} for details). This background can be reduced to $1.8 \pm 0.36$ events per day after applying the muon veto and IBD selection cuts. Fast neutrons produced by cosmic muons passing through the detector can reach the liquid scintillator without triggering the muon veto. The recoiling proton and the neutron capture can mimic the IBD signal. The fast neutron background is expected to be $0.01 \pm 0.01$, which is negligible.

The alpha particles emitted in decay chains of radioactive contaminants, $^{238}$U, $^{232}$Th and $^{210}$Po, can induce $^{13}{\rm C}(\alpha,n)^{16}{\rm O}$ reactions in the LS. The prompt signal produced by protons scattered off neutrons or the de-excitation of $^{16}$O and neutron capture can mimic IBD reactions. For the LS of the JUNO central detector, an initial purity level of of $10^{-15}$ g/g U/Th, $10^{-16}$ g/g K and $1.4\times 10^{-22}$ g/g $^{210}$Pb is estimated to be achievable without distillation~\cite{An:2015jdp}. Therefore, the $^{13}{\rm C}(\alpha,n)^{16}{\rm O}$ background rate is $0.05\pm 0.025$ per day based on the above assumption. Considering all the detector construction materials, the event rate of accidental coincidences of the non-correlated signals is estimated to be  $1.1 \pm 0.011$ in the fiducial volume~\cite{An:2015jdp}.

The event rates for those backgrounds are summarized in Tab.~\ref{tab:signalandbackground}. In addition, we show in Fig.~\ref{fig:neutrinospectrum} the spectra of reactor antineutrinos, other non-antineutrino backgrounds, and geo-neutrinos with the Th/U ratio fixed at the chondritic value.

\begin{center}
\footnotesize
\begin{tabular}{| p{3cm}| p{3cm}|}
\hline
    backgrounds & event rate/day \\
\hline
     $^{9}{\rm Li}-^{8}{\rm He} $ & 1.8  \\
\hline
     Fast neutrons & 0.01 \\
\hline
$^{13}{\rm C}(\alpha,n)^{16}{\rm O}$ & 0.05 \\
\hline
     Accidental events & 1.1  \\
\hline
\end{tabular}
\vspace{4mm}
\tabcaption{ \label{tab:nonantineutrinobackground} The non-antineutrino background event rate per day.}
\end{center}

\begin{center}
\includegraphics[width=7.5cm]{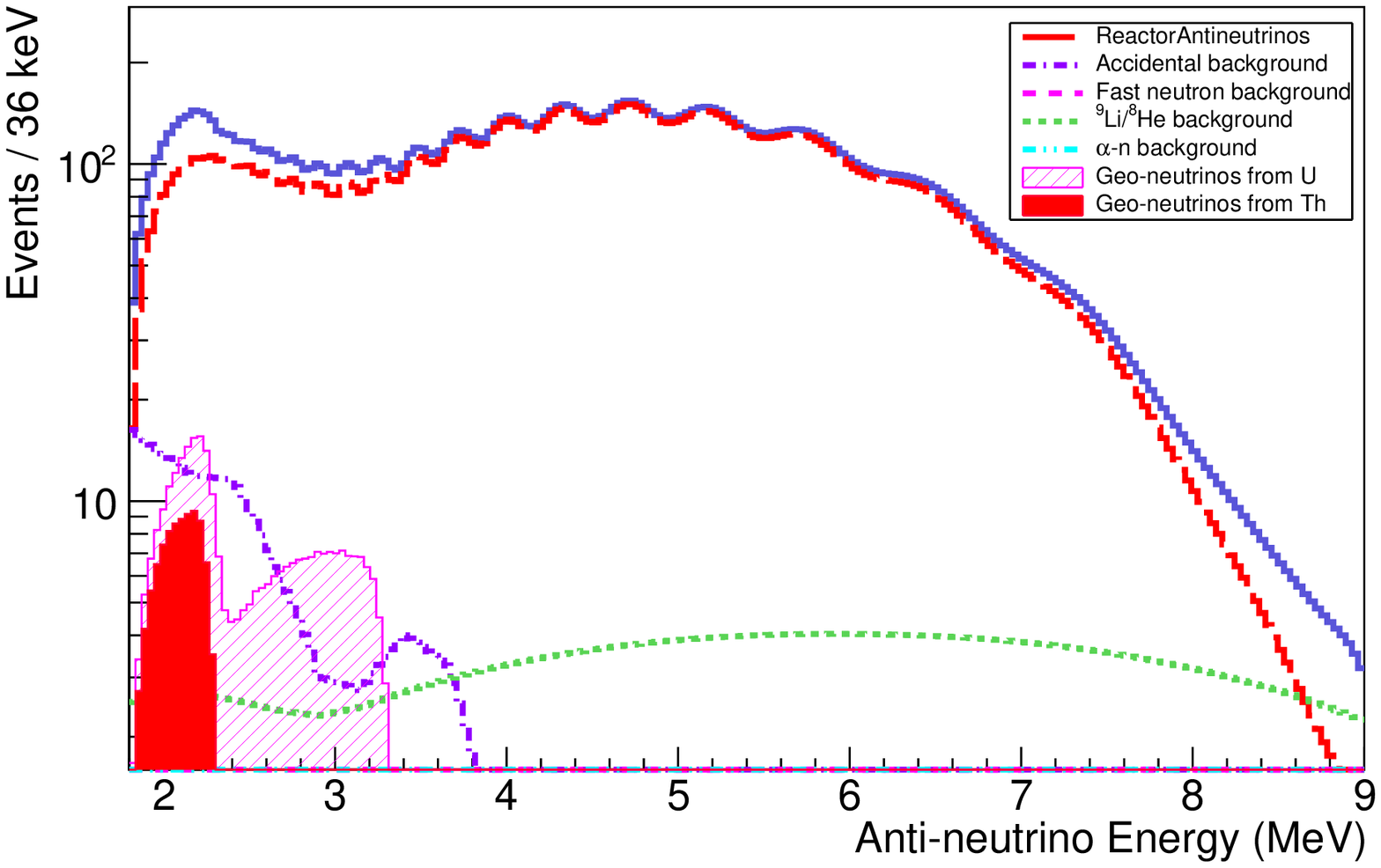}
\vspace{1mm}
\figcaption{\label{fig:neutrinospectrum}
The energy spectra of geo-neutrinos, reactor antineutrinos,
and other non-antineutrino backgrounds at JUNO for one year of data-taking.
The blue solid line is the total spectrum the red dashed line is the reactor antineutrinos.
The red solid area and pink area with parallel lines are antineutrinos from Th and U in the Earth, respectively. All the
non-antineutrino backgrounds are also shown, which can be directly read from the legend.}
\end{center}

\subsection{The signal to background ratio}

We can learn from Fig.~\ref{fig:neutrinospectrum} that the main background in the geo-neutrino energy range is the reactor antineutrinos
from the Yangjiang and Taishan nuclear power plants. The signal to background ratio (S/B) at different levels of the Yangjiang and Taishan thermal power is estimated.
From Fig.~\ref{fig:signaltobackground}, one can see that the signal to background ratio
is 46\% when the Yangjiang and Taishan nuclear power plants are totally switched off, but falls to 8\% when they are running at full power.
Without the reactor antineutrinos from Yangjiang and Taishan,
the main backgrounds are the non-neutrino backgrounds and other commercial reactor contributions.
\begin{center}
\includegraphics[width=7.5cm]{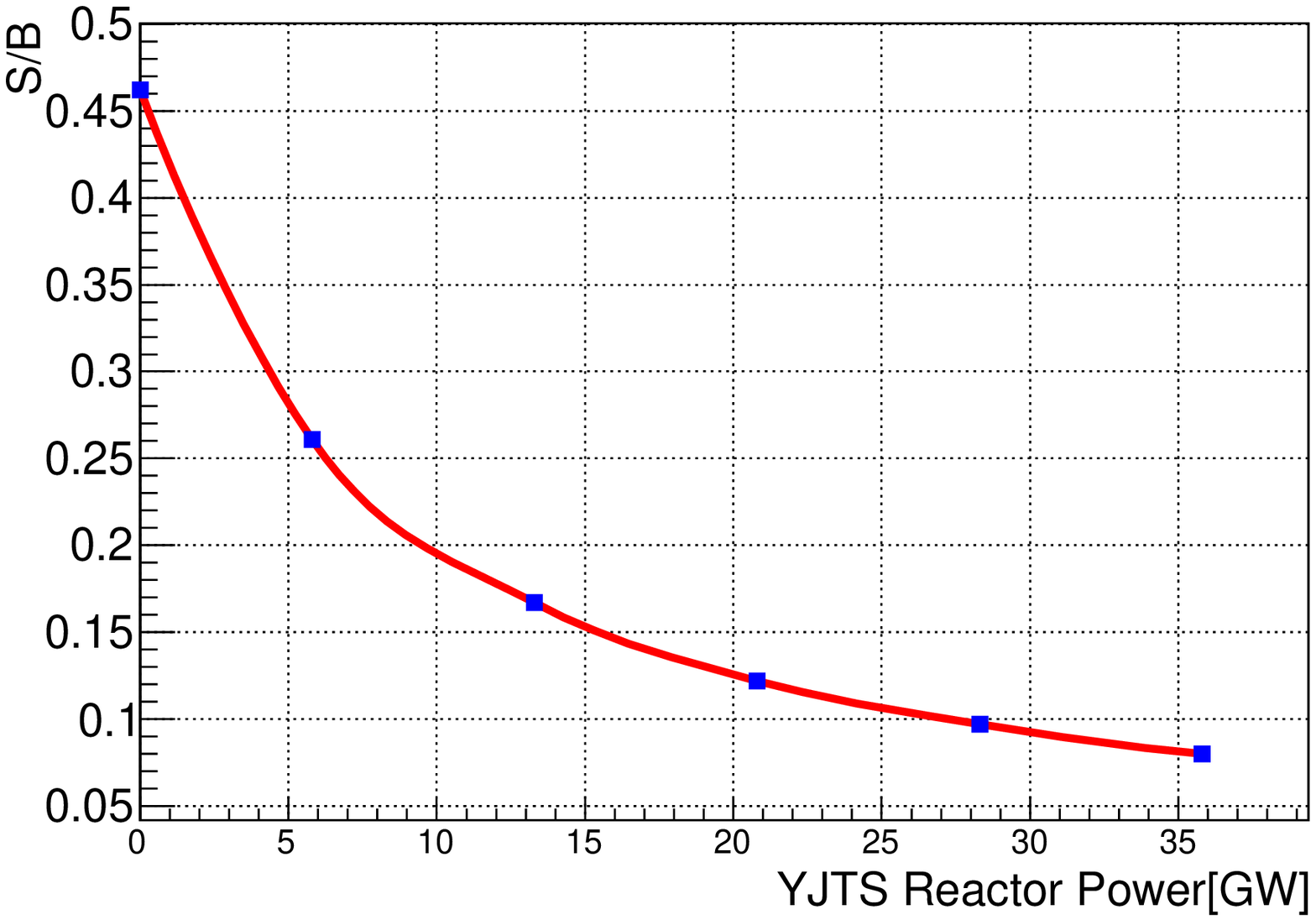}
\vspace{1mm}
\figcaption{\label{fig:signaltobackground}
Signal to background ratio (S/B) at different levels of Yangjiang (YJ) and Taishan (TS) running power.}
\end{center}

\section{Sensitivity study}
In order to extract the geo-neutrino signal from the high reactor antineutrino background, we employ the standard least-squares method to
quantitatively assess the potential of geo-neutrino measurements at JUNO. The predicted total antineutrino spectrum
(including both the signal and backgrounds) has been described in the previous section. A summary of the event numbers and corresponding rate and shape
systematic uncertainties for the signal and backgrounds is presented in Table~\ref{tab:signalandbackground}.
\begin{center}
\footnotesize
\begin{tabular}{| p{1.6cm} | p{1.8cm} | p{1.5cm} | p{1.7cm} | }
\hline
    source & events/year & rate uncertainty (\%)& shape uncertainty (\%)\\
\hline
     Geo-$\nu$s  & 408 (406) & NA & NA \\
     Reactor        & 16100 (3653) & 2.8 & 1 \\
     $^{9}{\rm Li}-^{8}{\rm He}$ & 657 (105) & 20 & 10 \\

     Fast $n$s & 36.5 (7.66) & 100 & 20 \\

     $^{13}{\rm C}(\alpha,n)^{16}{\rm O}$ & 18.2 (12.16) & 50 & 50 \\

     Accidental & 401 (348) & 1 & negl. \\
\hline
\end{tabular}
\vspace{2mm}
\tabcaption{ \label{tab:signalandbackground}
Event numbers and corresponding rate and shape systematic uncertainties of the signal and backgrounds used in the simulation.
In the second column, the first event numbers are for the energy range of [1.8, 9.0] MeV. The second event number
in the parentheses are for the energy range of [1.8, 3.3] MeV, where most of the geo-neutrino events are located.}
\end{center}

The $\chi^2$ function in the least-squares method is defined as follow:
\begin{eqnarray}
\label{eq:chi2}
\chi^2={\rm min}\left(\sum_{i=1}^{100}\frac{(N^{\rm obs}_{i}-N^{\rm pred}_{i})^2}{\sigma^2_{i,\rm stat}+\sigma^2_{i,\rm sys}}+\frac{\varepsilon^2_{\rm rea}}{\sigma^2_{\rm rea}}+\sum_{\rm ibg=1}^{4}\frac{\varepsilon^2_{\rm ibg}}{\sigma^2_{\rm ibg}}\right)\,,
\end{eqnarray}
where the index $i$ ($1\leq i\leq 100$) stands for the energy bin ranging from 1.8 MeV to 10 MeV.
$N^{\rm obs}_{i}$ is the total observed event number of the geo-neutrino signal, the reactor antineutrinos and other non-antineutrino backgrounds:
\begin{eqnarray}
N^{\rm obs}_{i}=N^{\rm obs}_{i,\rm geo}+N^{\rm obs}_{i,\rm rea}+\sum_{{\rm ibg}=1}^{4} N^{\rm obs}_{i,\rm ibg}\,,
\end{eqnarray}
where $N^{\rm obs}_{i,\rm geo}$, $N^{\rm obs}_{i,\rm rea}$ and $N^{\rm obs}_{i,\rm ibg}$ are calculated from the rates in
Table~\ref{tab:signalandbackground}, and spectra in Fig.~\ref{fig:neutrinospectrum} (i.e., Asimov data sets).
The statistical uncertainty in the $i$-th bin is defined as $\sigma^{2}_{i,\rm stat}=N^{\rm obs}_{i}$.
The uncorrelated systematic uncertainty in the $i$-th bin is calculated as
\begin{eqnarray}
\sigma^{2}_{i,\rm sys}=(N^{\rm obs}_{i,\rm rea}\cdot \sigma^{\rm shape}_{\rm rea})^{2} +
\sum_{{\rm ibg}=1}^{4} (N^{\rm obs}_{i,\rm ibg}\cdot \sigma^{\rm shape}_{\rm ibg})^{2}\,,
\end{eqnarray}
where $\sigma^{\rm shape}_{\rm rea}$ and $\sigma^{\rm shape}_{\rm ibg}$ are the relative shape uncertainties in Table~\ref{tab:signalandbackground}.
On the other hand, the rate uncertainties in Table~\ref{tab:signalandbackground} are included in Eq.~(\ref{eq:chi2}) by using the pull method, where
the prediction in Eq.~(\ref{eq:chi2}) is defined as
\begin{equation}
\label{eq:Npred}
N^{\rm pred}_{i}=N^{\rm obs}_{i,\rm rea}(1+\varepsilon_{\rm rea})+\sum_{{\rm ibg}=1}^{4}N^{\rm obs}_{i, \rm ibg}(1+\varepsilon_{\rm ibg})
+N^{\rm pred}_{i,\rm geo}\,,
\end{equation}
where $N^{\rm pred}_{i,\rm geo}$ will be specified in the following for different fitting scenarios.

\subsection{Scenario with a fixed Th/U ratio}

We first start with a fixed Th/U ratio at the chondritic proportion, in which we have
\begin{equation}
N^{\rm pred}_{i,\rm geo}=[N^{\rm obs}_{i,\rm geo}({\rm U})+N^{\rm obs}_{i,\rm geo}({\rm Th})]\times\alpha\,,
\end{equation}
where $N^{\rm obs}_{i,\rm geo}({\rm U})$ and $N^{\rm obs}_{i,\rm geo}({\rm Th})$ are the geo-neutrino events from U and Th respectively.
Therefore, we only have to determine a total geo-neutrino event normalization $\alpha$ in the fitting process. Considering the
increasing running time, we can determine the errors of geo-neutrino measurements in $\alpha$, which is shown in Fig.~\ref{fig:1sigmaerrorfixeduth}.
\begin{center}
\includegraphics[width=7.5cm]{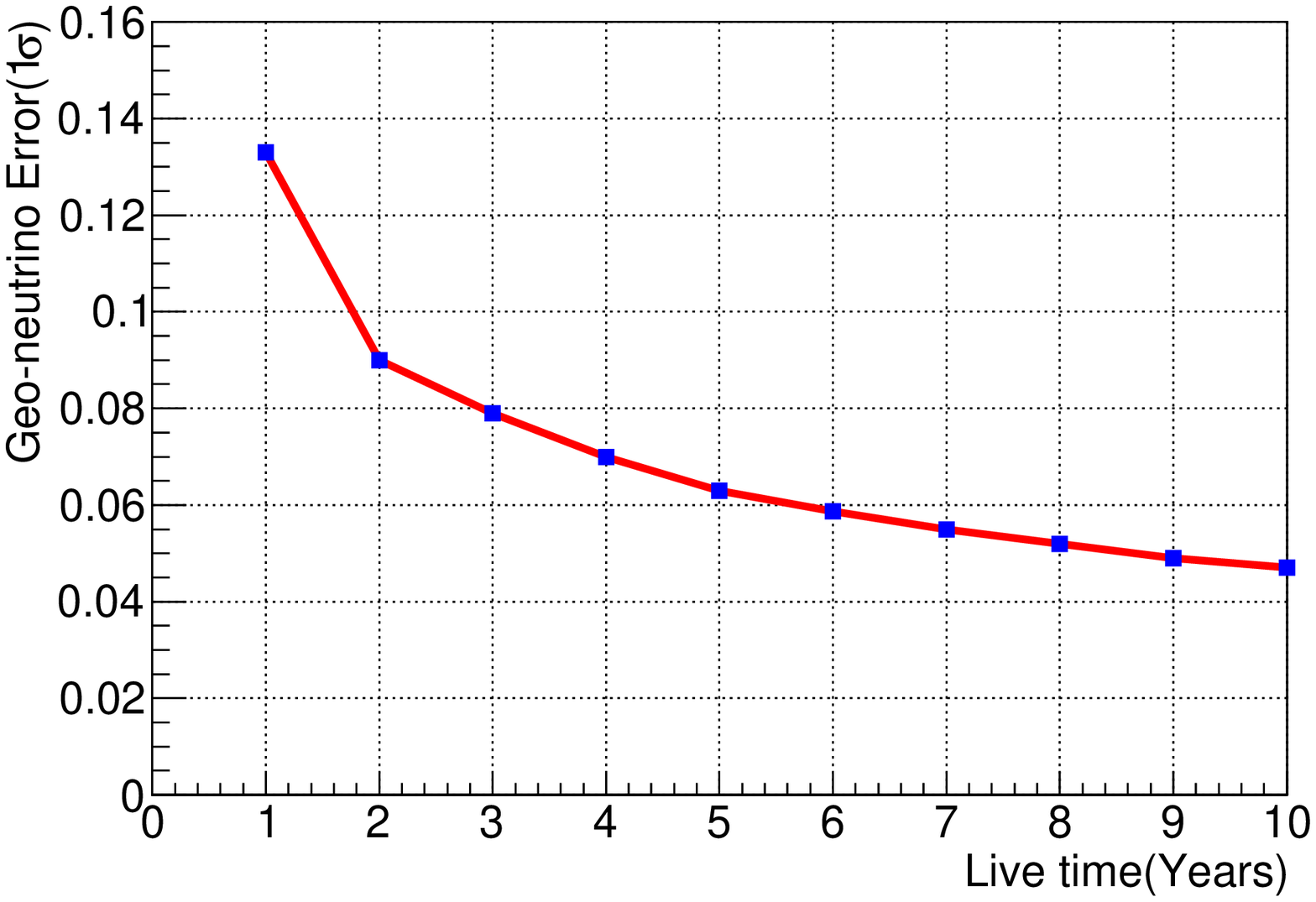}
\figcaption{\label{fig:1sigmaerrorfixeduth} The 1$\sigma$ uncertainty of geo-neutrino measurements as a function of running time at JUNO with a fixed chondritic Th/U ratio.}
\end{center}

With 1, 3, 5, and 10 years of data, the precision
of the geo-neutrino measurement with a fixed chondritic Th/U ratio is 13\%, 8\%, 6\% and 5\%, respectively,
which as expected, decreases with higher statistics.

\subsection{Scenario with a free Th/U ratio}

The high statistics geo-neutrino events at JUNO also provide us the potential to measure individually the U and Th contributions.
Therefore, in this senario, the experimental assumptions are the same as before. However,
two individual fitting parameters for the U and Th contributions of Eq.~(\ref{eq:Npred}) are assumed as follows:
\begin{equation}
N^{\rm pred}_{i,\rm geo}=N^{\rm obs}_{i,\rm geo}({\rm U})\times\alpha +N^{\rm obs}_{i,\rm geo}({\rm Th})\times\beta\,.
\end{equation}
Using the least-squares method, the two-dimensional $\chi^2$ distributions are shown in Fig.~\ref{fig:chi2uthfree}
for one (upper panel) and ten (lower panel) years of running, where the blue, green and red lines correspond to the allowed ranges of 1$\sigma$, 2$\sigma$ and 3$\sigma$
confidence levels respectively.
Within the first year of running, we derive the precision of Th and U contributions to be 80\% and 40\%, respectively.
With the increase of data-taking time, accuracy of 30\% and 15\% respectively can be obtained for ten years of running, which could
allow us to get high-significance measurements of the Th and U components in the Earth, and test the chondritic assumption of geological studies.
\begin{center}
\includegraphics[width=8cm]{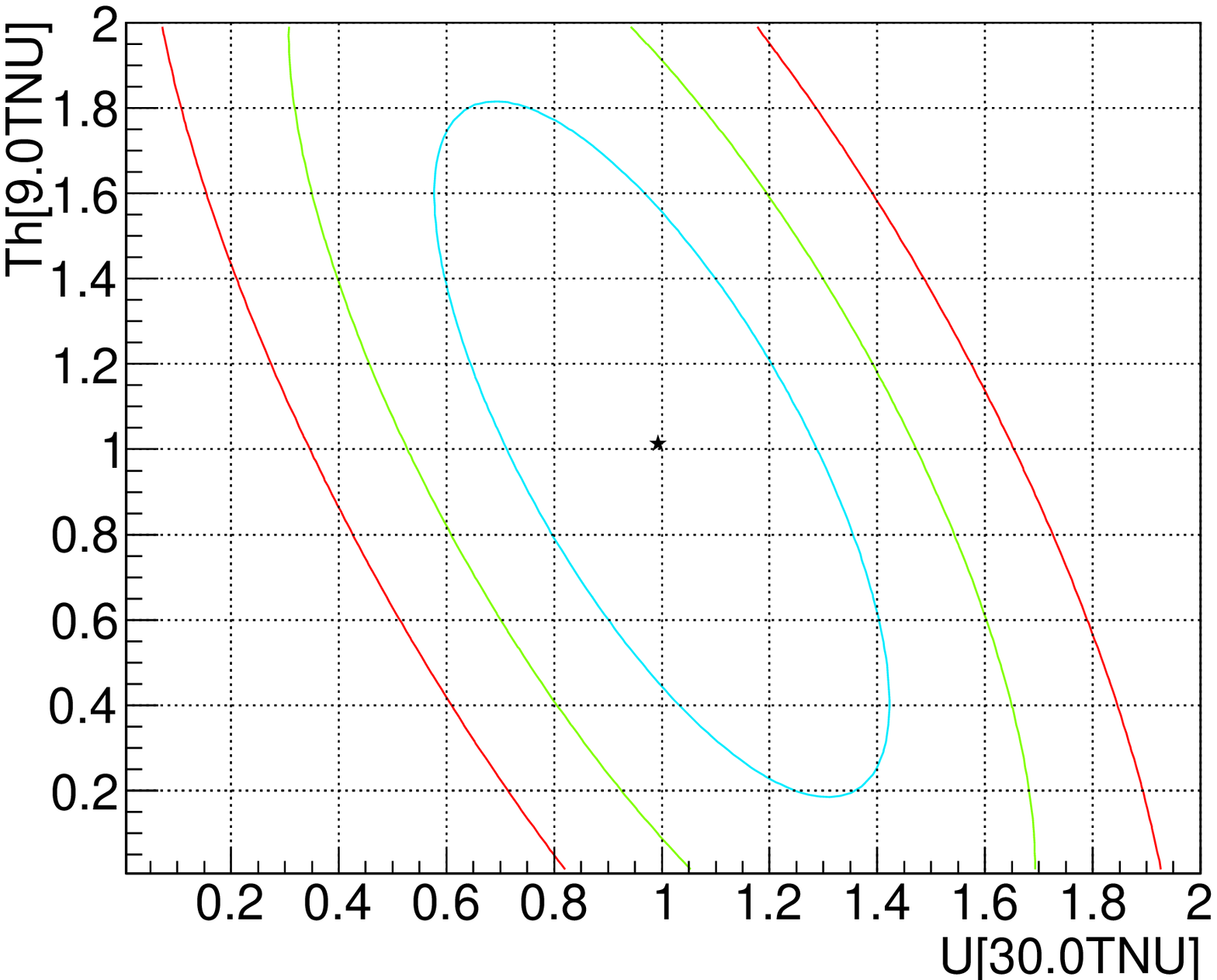}
\includegraphics[width=8cm]{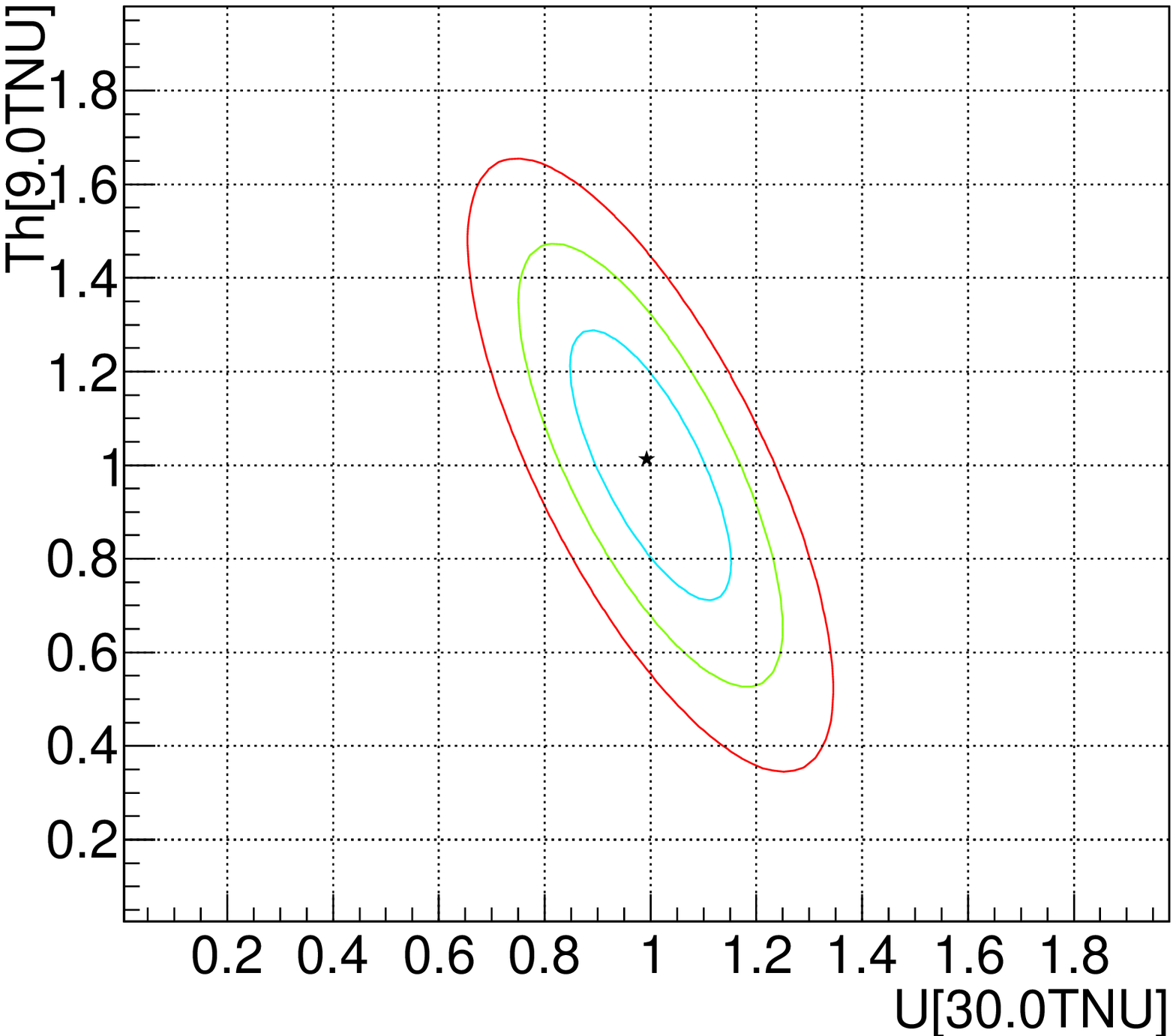}
\vspace{1mm}
\figcaption{\label{fig:chi2uthfree}
The $\chi^2$ distribution with a free Th/U ratio for one year (upper panel) and ten years (lower panel) running,
where the blue, green and red lines correspond to the allowed ranges of 1$\sigma$, 2$\sigma$ and 3$\sigma$
confidence levels respectively.}
\end{center}

\subsection{Extracting the mantle component}

Geo-neutrinos are generated from the crust and mantle regions of the Earth.
However, different from the crust, the mantle is almost unreachable and we have very limited knowledge of the abundance and distribution
of radioactive elements in the mantle.
The amount of radioactive heat coming from the mantle is unknown and model-dependent~\cite{Turcotte:2002,Anderson:2007,Allegre:1995,McDonough:1995,Lyubetskaya:2007,Javoy:2010}.

In principle the angular information of geo-neutrinos can help us to disentangle the mantle and crust contributions, but current (i.e., KamLAND and Borexino)
and next-generation (i.e., SNO+ and JUNO) experiments are using LS detectors, which are insensitive to the direction of low energy neutrinos. As a result,
we are left with an indirect substraction method of extracting the mantle component of geo-neutrino events. In this respect, we first have the experimental
measurement of the total geo-neutrino events $R({\rm total, exp.})$. If we can have an accurate prediction for the contribution of the crust $R({\rm Crust,theo.})$,
one can estimate the mantle component $R({\rm Mantle})$ as:
\begin{equation}
\label{eq:mantle}
R({\rm Mantle})=R({\rm total, exp.})-R({\rm Crust,pred.})\,,
\end{equation}
which tells us that the mantle measurement precision depends both on the precision of the total experimental measurement and on the
accuracy of the crust geo-neutrino prediction.

In the JUNO geo-neutrino prediction of Ref.~\cite{Strati:2014kaa}, an accuracy of 18\% is
estimated for the crust geo-neutrino contribution using the global reference Earth model in Ref.~\cite{Huang:2013}. Using this uncertainty value,
we obtain the blue solid line in Fig.~\ref{fig:geomantle}, which represents the $\chi^2$ distribution of mantle geo-neutrinos with a fixed chondritic Th/U ratio,
and gives $2\sigma$ separation of the mantle component. If one improve the uncertainty of geo-neutrino crust prediction to 8\%, which is the
level of KamLAND after a detailed local geological survey~\cite{Fiorentini:2012yk},
we can get the red dashed line in Fig.~\ref{fig:geomantle}, with a high-significance measurement
at an approximate 3.7$\sigma$ confidence level.
Comparing these two different assumptions, we can understand the importance of local geological studies in the vicinity of the experimental site.
\begin{center}
\includegraphics[width=8.5cm]{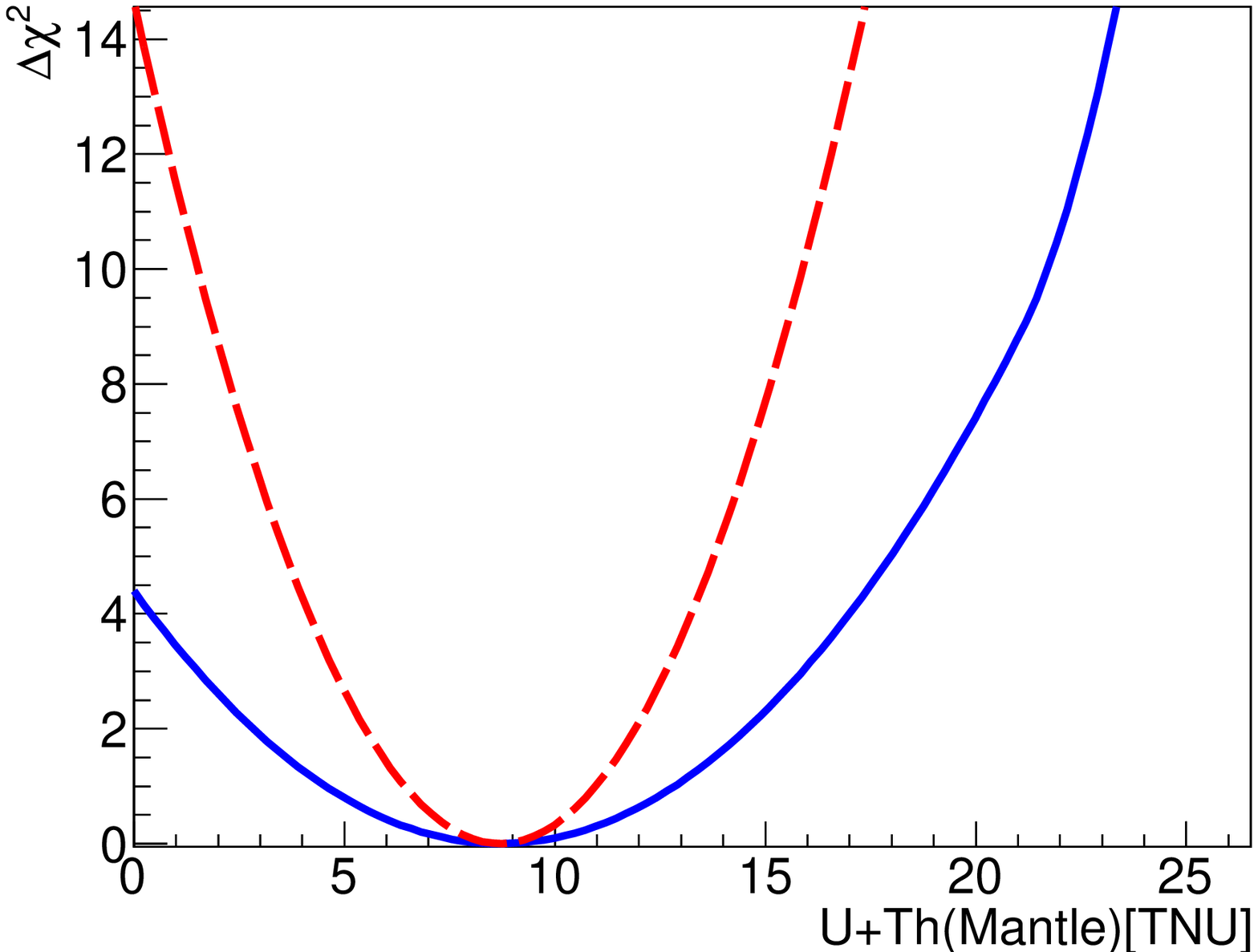}
\vspace{1mm}
\figcaption{\label{fig:geomantle}
The $\chi^2$ distribution of geo-neutrinos from mantle with a fixed chondritic Th/U ratio in the mantle. The blue solid line and red dashed line are for
the 18\% and 8\% precision of the crust geo-neutrino prediction, respectively.}
\end{center}

Besides the local geological studies, it is also important to combine mantle measurements at different locations,
which can help to distinguish the site-dependent crust components from the site-independent mantle contributions.
Therefore, we anticipate sizable improvement of the mantle geo-neutrino measurement by taking advantage of the combination of
the on-going KamLAND (Japan) and Borexino (Italy) experiments, and the SNO+ (Canada) and JUNO (China) experiments in the near future.

\section{Summary and future prospects}

With important geo-neutrino measurements at KamLAND and Borexino, the era of neutrino geoscience has arrived.
JUNO will join the family of geo-neutrino experiments, with its detector being at least 20 times larger than the existing detectors.
Within the first year of running, JUNO will record more geo-neutrino events than all other detectors will have accumulated to that time.
In this paper, we have presented the signal prediction and backgrounds for geo-neutrinos, and discussed the
precision level of geo-neutrino measurements at JUNO for the scenarios of the fixed and free chondritic Th/U ratios.
The possibility of extracting the mantle component is also discussed.

For future prospects, it would be important to reconstruct the angular information of geo-neutrinos,
which is important to separate the crust and mantle geo-neutrino components.
If $^6$Li or $^{10}$B can be doped in the current LAB recipe,
the reconstruction of geo-neutrino direction will be possible by measuring the displacement
between the positron and neutron event~\cite{Apollonio:1999jg,li6direction}.
Therefore, one can distinguish between the geo-neutrinos from the crust and mantle,
and meanwhile remove the reactor antineutrino background significantly.

The importance of collaboration between the particle physics and geological communities is emphasized.
To carry out a precision geo-neutrino measurement, a concentrated effort from the geological community, working in collaboration with particle physicists,
is necessary to acquire basic geological, geochemical and geophysical data for the regional area surrounding the detector.
For JUNO, we must develop a regional 3-dimensional model, which in practice is defined as the crust in the closest six $2^\circ \times 2^\circ$ crustal tiles and this critical data will need to be provided by a dedicated research effort. Experience tells us that in the continents the closest 500 km to the detector contributes half of the signal and it is this region that needs to be critically evaluated. This goal demands that the physical (density and structure) and chemical (abundance and distribution of Th and U) nature of the continent must be specified for the region.

\end{multicols}

\vspace{15mm}

\begin{multicols}{2}

\end{multicols}

\clearpage

\end{document}